\documentclass[%
 reprint,
 amsmath,amssymb,
 aps,
 prb,
]{revtex4-1}

\usepackage{graphicx}

\usepackage{dcolumn}
\usepackage{bm}

\begin{document}

\title{Superconductivity in Sm-doped 1,3,5-triphenylbenzene}

\author{Zheng Hu, Jin Si, Xiyu Zhu, Hai-Hu Wen}\email{hhwen@nju.edu.cn}

\affiliation{
National Laboratory of Solid State Microstructures and Department of Physics,
Collaborative Innovation Center of Advacnced Microstructures and , Nanjing University, Nanjing 210093, China
}

\date{\today}

\begin{abstract}
We report the discovery of superconductivity at about 4.3 K in the samarium doped 1,3,5-triphenylbenzene. By using a solid state reaction method,
the samarium doped 1,3,5-triphenylbenzene samples are successfully synthesized. These samples are characterized by magnetization, X-ray diffraction (XRD), scanning electron microscope (SEM) and energy dispersive spectroscopy (EDS) measurements. The X-ray diffractions reveal that the sample crystallizes in a space group of $P2/m$. The magnetization measurements reveal a superconducting transition at about 4.3 K. However,
the superconducting shielding fraction is only about 1\%, which is similar to the previous reports of the superconductivity in other aromatic hydrocarbons. Magnetization hysteresis loops (MHLs) of the sample show that it is a
typical type-II superconductor. Our result gives indications of possible superconductivity in this organic material.


\end{abstract}

\maketitle

\section{INTRODUCTION}
Organic superconductors have attracted a lot of attentions because of two fundamental reasons. Firstly, it was predicted that the superconducting transition temperature may reach a high value\cite{001}; secondly the pairing mechanism in many organic superconductors seems to be beyond the Bardeen-Cooper-Schrieffer (BCS) paradigm\cite{Wosnitza}. The first observation of superconductivity in organic materials can be traced back to 1980s, when Bechgaard et al. found superconductivity\cite{002} with a transition temperature ($T_c$) of 0.9 K under a pressure of 12 kbar in (TMTSF)$_2$PF$_6$. Here TMTSF represents tetramethyltetraselenafulvalene (C$_{10}$H$_{12}$Se$_{14}$) and is an electron donor molecule. Since then, researchers have found a series of superconductors of this type by replacing PF$_6$ with AsF$_6$, SbF$_6$, ClO$_4$, and so on\cite{003}. They are collectively referred to as the (TMTSF)$_2$X, where X represents the electron acceptor molecule, because they all have the same electron donor molecule TMTSF and the similar quasi-one-dimensional organic structures. Another major category of superconducting organic salts are (BEDT-TTF)$_2$X, where BEDT-TTF means bis(ethylenedithio)tetrathiafulvalene (C$_{10}$H$_8$S$_8$) and X can be I$_3$, Cu(SCN)$_2$, Cu[N(CN)$_2$]Br, Cu[N(CN)$_2$]Cl and so on. The highest superconducting transition temperature $T_c$ can be increased to 14.2 K in the form of $\beta$$^{'}$-(BEDT-TTF)$_2$ICl$_2$ by applying pressure to 8.2 GPa\cite{004}. Different from the former (TMTSF)$_2$X, (BEDT-TTF)$_2$X family is quasi-two-dimensional. There are abundant interesting physical phenomena in these systems, such as the competition between various ground states, including antiferromagnetic (AF) order and superconductivity\cite{005}.

Concerning the pairing mechanism of these organic superconductors, widely accepted picture is that the correlation effect may play a role here. For the two-dimensional organic superconductors, it is in the vicinity of Mott insulator. By tuning the ratio between the Coulomb interaction potential U and the band width W, one can change the system from a Mott insulator to superconducting state\cite{Wosnitza,Mott1,Mott2}. In 1982, experimental data of specific-heat on (TMTSF)$_2$ClO$_4$ was explained by the BCS theory with strong superconducting fluctuations due to the low dimensionality\cite{026}. In 2004, study of the compounds (TMTSF)$_2$(ClO$_4$)$_{1-x}$(ReO$_4$)$_x$ showed the suppression on superconductivty effect by nonmagnetic impurities\cite{027}, which suggests that the superconducting order parameter may have a sign change with momentum. Furthermore, the study of the spin lattice relaxation rate($1/T_1$) suggests existence of anisotropic order parameter with line nodes on the Fermi surface\cite{028}. This is against the original electron-phonon coupling based BCS picture. However, thermal conductivity experiments shows the opposite result, it indicates nodeless superconducting gap function in the same organic superconductor\cite{029}. Some theoretical investigations show that in the quasi-one-dimensional system (TMTSF)$_2$PF$_6$, superconductivity possibly has a p-wave pairing symmetry with spin triplet pairing state\cite{006}, while in the quasi-two-dimensional system (BEDT-TTF)$_2$X, it could be the d-wave pairing symmetry with spin singlet state\cite{007}. In 2012, a refined field-angle-resolved calorimetry measurement on (TMTSF)$_2$ClO$_4$ supports d-wave singlet pairing\cite{030}. However, one year later, experiment on (TMTSF)$_2$ClO$_4$ by muon-spin rotation showed no indication of gap nodes on the Fermi surface and suggested p-wave triplet pairing\cite{031}. Nuclear magnetic resonance (NMR) experiments on $\kappa$-(ET)$_2$Cu[N(CN)$_2$]Br exclude the BCS electron-phonon mechanism and suggest an unconventional pairing state with possible nodes in the gap function\cite{032,033,034}. Recent NMR experiments on $\beta$-(ET)$_2$SF$_5$CH$_2$CF$_2$SO$_3$ also supports the pair symmetry of d-wave and suggests that pairing in this compound is driven by charge fluctuations\cite{035}. In addition to NMR experiments, angle-resolved thermal-conductivity measurements on $\kappa$-(ET)$_2$Cu(NCS)$_2$ is in consistent with d-wave pairing\cite{036}. Numerous scanning tunneling spectroscopy measurements on $\kappa$-(ET)$_2$X have been done, the tunneling conductance curve obtained at the conducting plane and the temperature dependence of tunneling spectra all agreed on d-wave pairing\cite{037,038,039,040,041,042}. In contrast to this conclusion, most experiments of specific heat measurements show however the feature of nodeless gaps\cite{043,044,045,046,047}, this gives puzzles in this field. Thus some debates still exist about the pairing symmetry of superconducting order parameter of the organic superconductors, but many more experiments show the existence of gap nodes, which excludes the possibility of phonon mediated pairing, but rather by other medium, like spin or charge fluctuations.

Beside these organic charge-transfer salts, superconductivity has also been discovered in many other carbon based materials, such as graphite, molecular crystal of C$_{60}$ and other aromatic materials. Back to 1965, researchers first discovered superconductivity in carbon based compounds\cite{008}. Superconductivity was induced by intercalating alkali-metal potassium into graphite. Along this direction, in 1991, researchers found superconductivity with a transition temperature of 18 K in potassium doped C$_{60}$\cite{009}. Later it was found that $T_c$ of Cs$_2$RbC$_{60}$ can reach 33 K, and for Cs$_3$C$_{60}$ the $T_c$ is 40 K at 15 kbar\cite{010,011}. Since the spin relaxation rate 1/T$_{1}$ of Rb$_3$C$_{60}$ shows a clear peak at T$_c$ and the consistency with the Hebel-Slichter theory, and the gap ratio $\Delta/k_BT_c\approx$ 3.6, both are consistent with the conventional BCS theoretical predictions, thus the superconductivity may be attributed to the phonon-based mechanism\cite{012}. Further theoretical calculations reveal that, albeit the large Coulomb interaction and presence of narrow band in A$_3$C$_{60}$ superconductors, the system is still not considered to be a Mott-Hubbard insulator. The superconductivity can be explained by a strong coupling picture with an enhanced Coulomb pseudopotential $\mu\ast$ due to the strong electron interactions\cite{013}.

Recently, aromatic hydrocarbons have attracted a lot of attention because some of them can show superconductivity by doping alkali-metal or alkali-earth-metal. In 2010, researchers synthesized K$_{3.3}$picene and found superconductivity with $T_c$ of 7 K and 18 K\cite{014}. Subsequently the superconductivity with 5 K in potassium-doped phenanthrene and superconductivity with 33 K in potassium-doped 1,2:8,9-dibenzopentacene were reported\cite{015,016}. Unfortunately, no repeating experiment about the superconductivity at 33 K has been reported. Some experiments show that superconductivity may disappear if the aromatic materials are made in a pure state\cite{048,049}. While it remains unclear whether the compounds showing superconductivity are the same as those materials without superconductivity\cite{048,049}. There is a theoretical prediction that $T_c$ is directly proportional to the number of benzene rings\cite{016} in the constructing molecule. In recent years, it was reported that there might be very high superconducting transition temperature in K$_3$p-terphenyl, whose $T_c$ can be as high as 120 K\cite{017}. However, the diamagnetic volume at low temperatures in K$_3$p-terphenyl is only about 0.04\%, thus it is insufficient to conclude that diamagnetic behavior is derived from superconductivity\cite{018}. These experiments suggest that researchers can obtain superconductors with different $T_c$ by adjusting the arrangement and number of benzene rings in molecular crystals. The existence of positive pressure dependence of $T_c$ and its correlation with enhanced local magnetic moments suggest that the superconductivity may be unconventional\cite{015,019}. Since the superconducting shielding volume in all these aromatic superconductors is still low, it remains unclear what is the actual crystal structure and chemical formula which is responsible for superconductivity. Considering that some of the reports of superconductivity in these materials are hard to be reproduced, one cannot rule out the possibility that some observations of superconductivity are due to impurity phases.

In this paper, we report the synthesis of a new aromatic material which shows indication of superconductivity by doping samarium, a magnetic rare-earth metal, into 1,3,5-triphenylbenzene with molar ratio of 3 : 1. The $T_c$ of Sm$_3$1,3,5-triphenylbenzene is about 4.3 K and the magnetic shielding fraction is only about 1\%. This kind of small shielding fraction seems to be a common feature in aromatic hydrocarbons\cite{020}. Magnetization measurements show that the superconductivity observed in this compound has the characteristics of a type-II superconductor.

\section{EXPERIMENT DETAILS}

The samples are synthesized by means of solid state reaction. Samarium metal ($>$99\%, GRINM) is ground into powder and mixed with 1,3,5-triphenylbenzene($>$99\%, Alfa Aesar) in a molar ratio 3:1. Then, the mixture is ground and pressed into pellets. All procedures are handled in a glove box filled with argon gas (O$_2$ and H$_2$O are less than 0.1ppm). The pellet is then put into an Al$_2$O$_3$ crucible and sealed in a quartz tube under a high vacuum. The quartz tube is put into furnace and heated up to 500 K in 600 minutes, and stayed at this temperature for 6 days. Finally we get samples with dark color. The samples are not sensitive in air, thus we can make several kinds of measurements. The X-ray diffraction (XRD) measurements are performed on the raw material of 1,3,5-triphenylbenzene and the resultant sample at room temperature with a Bruker D8 Advanced diffractometer with the CuK$_\alpha$$_1$ radiation. The XRD patterns are obtained in the 2$\theta$ range from 10$^{\circ}$ to 90$^{\circ}$ with a scanning rate of 0.2 sec/step. The morphology and surface compositions of Sm$_3$1,3,5-triphenylbenzene are investigated by scanning electron microscope (SEM) and energy dispersive spectroscopy (EDS) methods with the instrument of Phenom ProX (Phenom). An accelerating voltage of 15 kV is used for the EDS measurements. The magnetization measurements were carried out on a superconducting quantum interference device with the vibrating sample option (SQUID-VSM, Quantum Design).

\section{RESULTS AND DISCUSSION}

\begin{figure}
  \includegraphics[width=8.5cm]{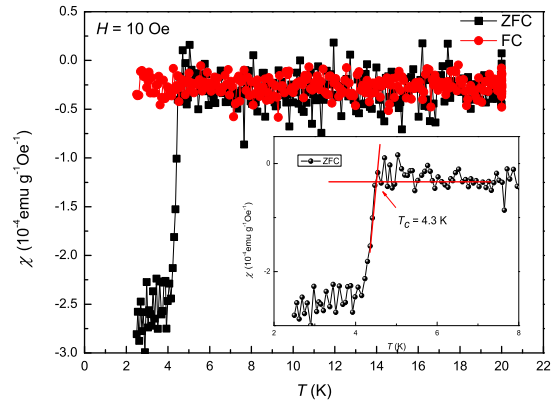}
\caption{Temperature dependence of magnetic susceptibility $\chi$ measured in ZFC and FC modes at 10 Oe for the sample Sm$_3$1,3,5-triphenylbenzene. Inset presents the enlarged view of $\chi$ versus T near the transition temperature.}
 \label{fig1}
\end{figure}

Figure~\ref{fig1} shows the temperature dependence of magnetic susceptibility $\chi$ for a bulk sample of Sm$_3$1,3,5-triphenylbenzene measured in zero-field-cooled (ZFC) and field-cooled (FC) modes under a magnetic field of 10 Oe. The sample has a mass of about 1 mg. Since the total magnetization signal is very small, the data is a bit noisy. When the field is increased, the total magnetization becomes stronger and noise is weaker. The $\chi$ versus T in the inset shows a drastic decrease below 4.3 K which is determined as the superconducting transition temperature $T_c$. We can see that the transition is very sharp with width less than 1 K. The diamagnetic signal $\chi$ measured in the ZFC mode can be attributed to the magnetic shielding of the Meissner effect in the superconducting state. By subtracting ZFC data from FC data at 2.5 K, we can obtain $\Delta$$\chi$, which is about $2.5\times10^{-4}$ emu g$^{-1}$ Oe$^{-1}$. Assuming that the density of the sample is about 3.5 g/cm$^{3}$, according to formula $V_S = 4\pi\Delta\chi\rho$ (V$_S$ means shielding fraction), we can get a  magnetic shielding fraction of about 1\%. It should be emphasized that, for all the polycyclic aromatic hydrocarbon (PAH) superconductors reported to date, the shielding fractions are all very small\cite{020}. Although these shielding fractions can be increased by pressing the powder samples into tablets\cite{014,015}, it is still very small compared to other superconductors. It has been discussed that these small shielding fractions in powder samples may be attributed to the penetration depth effect. According to London's theory, when the penetration depth is larger than the grain size, the measured diamagnetization can be very small\cite{014}. For organic superconductors, the superfluid density is very low, thus the London penetration depth is very large, yielding a possibility for this explanation. But this picture holds valid only when the sample is constructed by unconnected grains and superconducting phase is in the form of percolative.  Another characteristic of the magnetic susceptibility is that, the signal measured in the FC mode is small or unmeasurable. The small FC magnetization and the clear difference of the ZFC and FC magnetizations were observed in other type-II superconductors with strong vortex pinning\cite{VortexBaK122}. Following this logic, the small FC magnetization in present samples may be attributed to the vortex pinning, perhaps arising from the pinning by the grain boundaries. In general, the magnetic shielding volume estimated in London's theory in this type of superconductors is small, which needs further investigation to unravel the reason.

In order to rule out other possibilities for the observed superconductivity, we have checked to literatures and found that some Sm-based superconductors exist. These include for example Sm$_1$phenanthrene (T$_c$ = 6 K)\cite{024}, Sm$_1$picene (T$_c$ = 4 K)\cite{025}, Sm$_1$chrysene (T$_c$ = 5-6 K)\cite{025}, all show small superconducting volume. For inorganic superconductors, we know that SmBa$_2$Cu$_3$O$_7$, and SmFeAsO$_{1-x}$F$_x$ are superconductors. By checking to literatures, we have not found any form of superconductors made by the alloy or compounds of Sm and carbon, either in binary or multi-element compounds. Besides, we have tried to synthesize the Sm with other organic molecules, such as Sm-doped 1,3,5-Tris(bromomethyl)benzene, and Sm-doped triphenylene with the same method, the samples are not superconductive. Furthermore, if we synthesize the sample at an elevated temperature, such as 600 K with the mixture of Sm and 1,3,5-triphenylbenzene, no uniform phase is formed and no superconductivity is observed on those samples. All these point to the fact that the superconductivity observed here may arise from the Sm-doped 1,3,5-triphenylbenzene.

\begin{figure}
  \includegraphics[width=8.5cm]{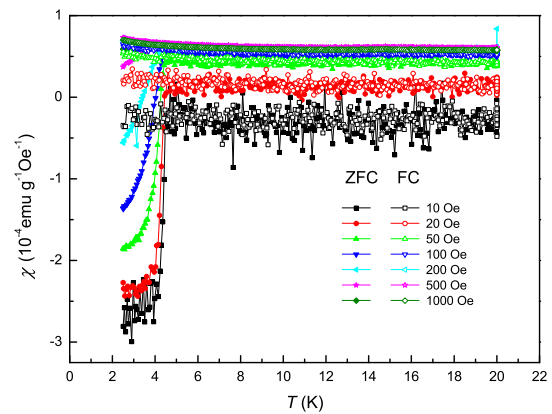}
\caption{Temperature dependence of magnetic susceptibility $\chi$ measured in ZFC and FC modes under different magnetic fields.}
 \label{fig2}
\end{figure}

Figure~\ref{fig2} shows the temperature dependence of magnetic susceptibility $\chi$ under different magnetic fields with measurements in ZFC and FC modes for the sample Sm$_3$1,3,5-triphenylbenzene. As we can see, the diamagnetic signals get smaller and smaller with increasing magnetic fields. However, even the magnetic field is increased up to 1000 Oe, we can still see the drop of $\chi$ at 4 K. This result suggests that the H$_{c2}$ of the sample may be large and far beyond 1000 Oe. It should be noted that even though the step-like transition still exists, above 200 Oe, the diamagnetic signal disappears, instead a paramagnetic background emerges. This may be understood in the way that the superconducting part inside the sample is very small, most areas of the sample are non-superconducting part which contributes a relatively large paramagnetic signal. Due to the paramagnetic background, the magnetic susceptibility of the non-superconducting part becomes more and more prominent when the magnetic field is getting higher and higher. Such clear paramagnetic background signals can be clearly seen in the following measurements of magnetization hysteresis loops (MHLs). One can see from Figure~\ref{fig2} that, when the magnetic field is small, the signal shows some noise. This is due to the small superconducting signal of the sample compared with the resolution of the instrument.

\begin{figure}
  \includegraphics[width=8.5cm]{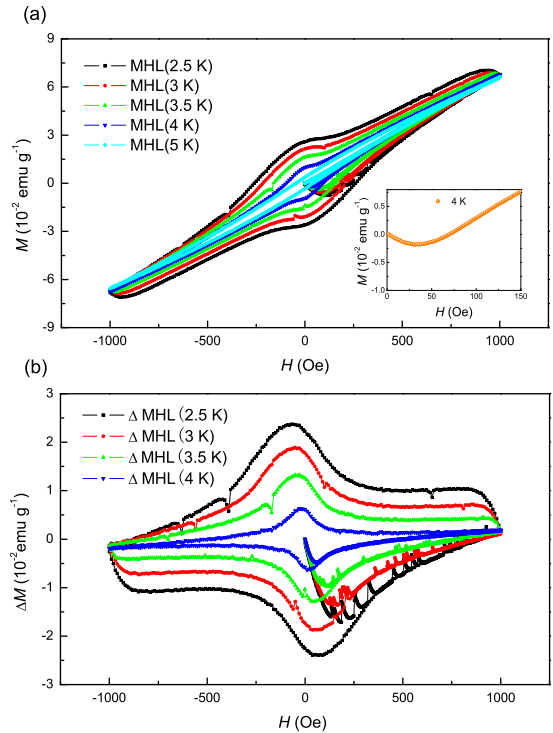}
\caption{(a) magnetization hysteresis loops (MHLs) of Sm$_3$1,3,5-triphenylbenzene from 2.5 K to 5 K. Inset in lower right shows the enlarged view of the MHLs at 4 K. (b) Magnetization hysteresis loops ($\Delta$MHLs) of Sm$_3$1,3,5-triphenylbenzene subtracted the MHL measured at 5 K, the latter is taken as the background.}
 \label{fig3}
\end{figure}
Figure~\ref{fig3}(a) shows the magnetization hysteresis loops (MHLs) of Sm$_3$1,3,5-triphenylbenzene with the range of magnetic field from -1000 Oe to 1000 Oe at different temperatures. The enlarged view of MHL at 4 K is shown in inset of Figure~\ref{fig3}(a). As we can see, the Meissner effect-like behavior is detectable, although a relatively strong paramagnetic background exists. From the Meissner effect-like behavior, we can roughly determine the H$_{c1}$ of the sample at 2.5 K, which is about 89 Oe. This is smaller than the value of 175 Oe at 2 K in K$_3$phenanthrene with a T$_c$ of 5 K\cite{015}. Here H$_{c1}$ means the threshold at which magnetic field starts to penetrate into the sample and is determined by the point of deviation from the linear line on the slope of the initial magnetization curve\cite{021}. The inset in Fig. 6 shows the specific way that how to define H$_{c1}$. The shielding behaviors of MHLs decay gradually with increasing temperature and disappear at 5 K. It should be emphasized that when temperature is lower than 4 K, the hysteresis loop exhibits some jumps between 0 Oe and 1000 Oe. This phenomenon appears in repeated measurements, which we think may be induced by the flux jump effect. Figure~\ref{fig3}(b) shows the subtracted magnetization hysteresis loops ($\Delta$MHLs) of Sm$_3$1,3,5-triphenylbenzene with MHL at 5 K as background, i.e., $\Delta M = M(H, T)- M(H, 5K)$. And the Meissner effect-like behaviors become more visible after subtracting the background. The superconductivity can be repeatedly observed with samples in different rounds of synthesis. We have tried 10 rounds of synthesis, at least 50\% reproducibility is achieved if the same fabrication procedures are followed.

\begin{figure}
  \includegraphics[width=8.5cm]{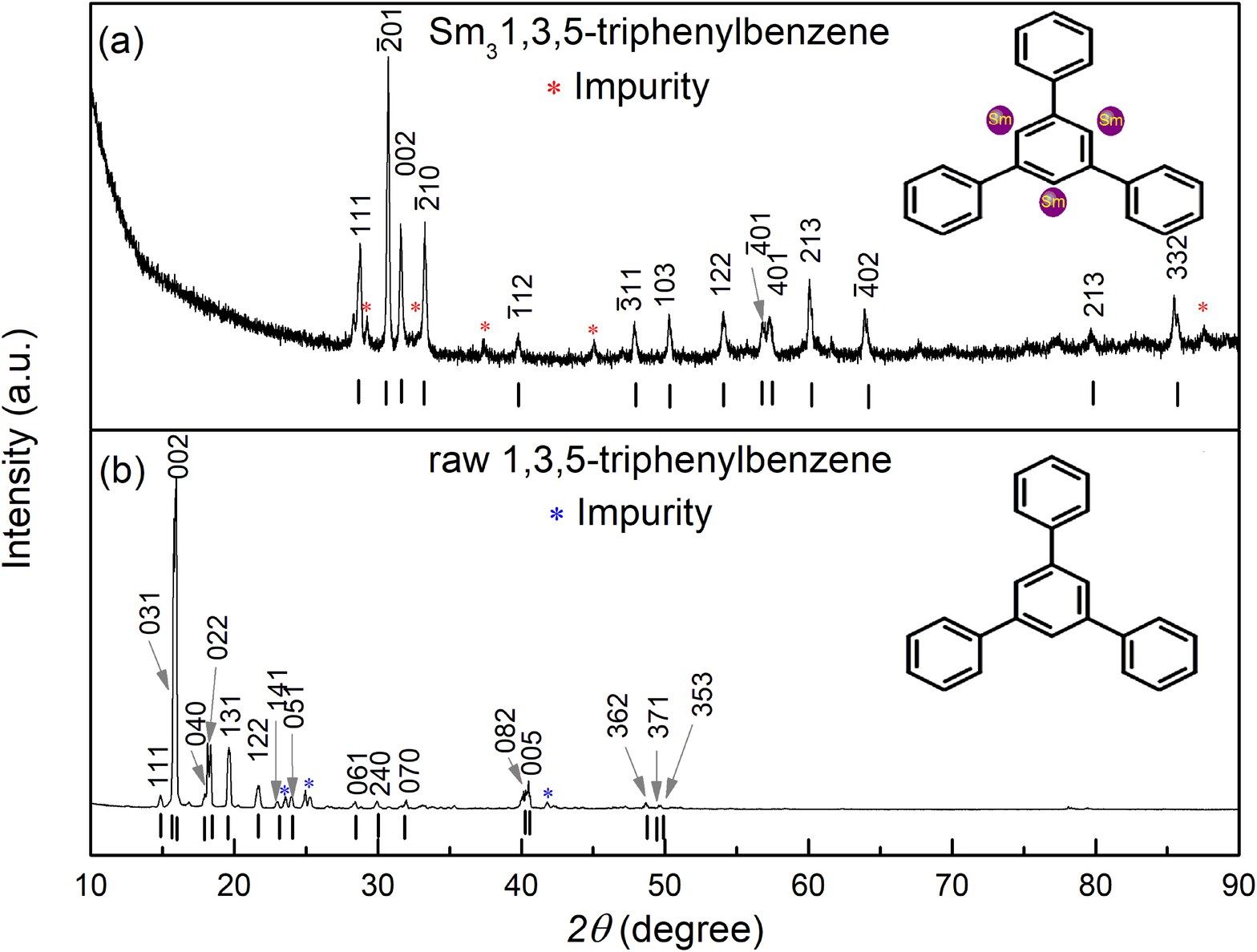}
\caption{(a) X-ray diffraction patterns for the Sm$_3$1,3,5-triphenylbenzene powders. Upper right inset in (a) presents the schematic molecular structure of Sm$_3$1,3,5-triphenylbenzene. (b) X-ray diffraction patterns for the raw material 1,3,5-triphenylbenzene. Upper right inset in (b) presents the schematic molecular structure of 1,3,5-triphenylbenzene.}
 \label{fig4}
\end{figure}

Figure~\ref{fig4}(a) shows the x-ray diffraction(XRD) patterns of the samarium doped 1,3,5-triphenylbenzene with molar ratio 3:1. Figure~\ref{fig4}(b) shows the XRD pattern of the raw material of 1,3,5-triphenylbenzene, which crystallizes in a space group of $Pmmm$. The lattice parameters determined here are $a = 7.47\mathring{A}$, $b = 19.66\mathring{A}$, $c = 11.19\mathring{A}$, being consistent with the results reported before\cite{022}. Comparing the XRD results in Figure~\ref{fig4}(a) and (b), it is obvious that the XRD pattern changes a lot after doping samarium into 1,3,5-triphenylbenzene. By fitting to the XRD pattern, we find that the Sm-doped sample may crystallize in a space group of $P2/m$. And the lattice parameters determined are $a = 6.7384\mathring{A}$, $b = 4.4798\mathring{A}$, $c = 5.6726\mathring{A}$ and $\beta = 90.746^{\circ}$ for Sm$_3$1,3,5-triphenylbenzene. The space group and lattice constants are determined by the FULLPROF program with a self-consistent fitting to the peak positions\cite{023}. Upper right insets in both figures present the schematic molecular structure of Sm-doped and pure 1,3,5-triphenylbenzene. For 1,3,5-triphenylbenzene, which consists of four benzene rings, three benzene rings are connected with the middle benzene ring by C-C bonds at the interval position. We want to stress that, for Sm$_3$1,3,5-triphenylbenzene, drawn in the inset of Fig.~\ref{fig4}(a) is just a schematic one. From the XRD data we can only get the message of space group and approximate lattice structure. Here the difficulty is that we don't know the occupation sites of the Sm atoms and the 1,3,5-triphenylbenzene molecules as well as the orientations of the latter within an unit cell. For a precise determination of the structure we need to do further refined diffraction measurements with synchrotron experiment and the quantum theory calculations. These are left for a future effort. However, since the superconducting shielding fraction seems to be small, it remains unclear whether the structure determined here reflects the superconducting phase or not. One possibility for the small magnetic shielding volume is that the penetration depth is much larger than the superconducting grain size, thus the total magnetization of the granular samples can be very small.

\begin{figure}
  \includegraphics[width=8.5cm]{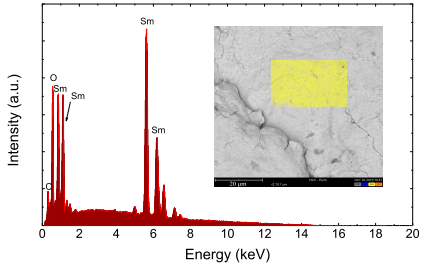}
\caption{Scanning electron microscope image and compositional analysis of the Sm$_3$1,3,5-triphenylbenzene sample. The main panel shows the EDS spectrum from which samarium, Carbon and Oxygen are detectable. The inset shows the SEM image of one area of the sample. The rectangular frame with yellow color shows the area for the EDS measurement. The brightness of the frame reflects the composition of Sm.}
 \label{fig5}
\end{figure}

Figure~\ref{fig5} shows the scanning electron microscope (SEM) image and composition analysis. The inset displays the SEM photograph of the sample surface and the yellow region shows the distributions of Sm. The brightness of yellow color gives the intensity of the EDS signal of samarium. One can see clearly that samarium is uniformly distributed in the body of the sample. The main panel represents the compositional analysis (EDS) spectrum with the indices of different elements. The position and intensity of the peaks in the EDS spectrum correspond well to the expected elements. The oxygen element peaks can also be seen due to exposure to air.

\begin{figure}
  \includegraphics[width=8.5cm]{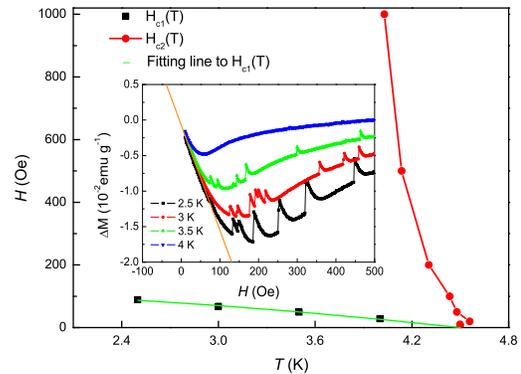}
\caption{The $H$ versus $T$ phase diagram. The black squares represent $H_{c1}$ and red circles represent $H_{c2}$.
Middle inset presents the way of defining the $H_{c1}$.}
 \label{fig6}
\end{figure}
Figure~\ref{fig6} shows the $H$ versus $T$ phase diagram of Sm$_3$1,3,5-triphenylbenzene sample. On the basis of data shown in the middle inset, we obtain $H_{c1}$ versus $T$, which is represented by black squares. The way of defining the $H_{c1}$ has been given when discussing about   Fig.~\ref{fig3}. We try to fit $H_{c1}$(T) with the empirical formula $H_{c1}(T) = H_{c1}(0)[1-(T/T_c)^2]$. We get the value of $H_{c1}\approx$ 126 Oe at 0 K. The green line shows the fitting line. Based on the data shown in Fig.~\ref{fig2}, the $H_{c2}$ versus $T$ phase line is obtained and drawn by red circles. Here $T_c$ is determined by the crossing point of the two linear lines corresponding to the normal state flat background and the steep superconducting transition line, as shown in the inset of Fig.~\ref{fig1}. It is difficult to determine the value of $H_{c2}$ at 0 K. However, based on the trend of $H_{c2}$(T) near T$_c$, we can conclude that the $H_{c2}(0)$ should be very high.

\section{CONCLUSIONS}
In conclusion, by using a solid state reaction method, we successfully synthesize the Sm$_3$1,3,5-triphenylbenzene samples. The magnetic susceptibility of the sample shows a diamagnetic transition at about 4.3 K. The diamagnetic transition is proved to be a superconducting transition by further measuring the temperature dependence of magnetization under different magnetic fields and MHLs. The MHLs indicate a type-II superconductivity for this compound. Fitting to the index peaks of XRD patterns of the sample Sm$_3$1,3,5-triphenylbenzene tells that the it crystallizes in a space group of $P2/m$ with $a = 6.7384\mathring{A}$, $b = 4.4798\mathring{A}$, $c = 5.6726\mathring{A}$ and $\beta = 90.746^{\circ}$. The SEM image shows that samarium element uniformly distributes in the sample. Since the superconducting volume determined here is still quite small, it remains to know whether the superconductivity can be attributed to the determined structure. Thus it is worthwhile to have further efforts to resolve the superconducting phase. However, we need to notify that, Sm doping is necessary for the emergence of superconductivity.

\section{ACKNOWLEDGMENTS}
We appreciate the useful discussions with A. V. Balatski. This work was supported by National Key R\&D Program of China (grant number: 2016YFA0300401, 2018YFA0704200), National Natural Science Foundation of China (grant number: 11534005) and the Strategic Priority Research Program of Chinese Academy of Sciences (grant number: XDB25000000).


\begin{thebibliography}{00}
\bibitem{001}W. A. Little, Phys. Rev. \textbf{134}, A1416 (1964).
\bibitem{Wosnitza}For a recent review, see for example, A. Ardavan, S. Brown, S. Kagoshima, K. Kanoda, K. Huroki, H. Mori, M. Ogata, S. Uji, and J. Wosnitza. J. Phys. Soc. Japan 81, 011004(2012).
\bibitem{002}D. J\'erome, A. Mazaud, M. Ribault, and K. Bechgaard, J. Phys. Lett. \textbf{41}, 95 (1980).
\bibitem{003}S. S. P. Parkin, M. Ribault, D. J\'erome, and K. Bechgaard, J. Phys. C \textbf{14}, 5305  (1981).
\bibitem{004}H. Taniguchi, M. Miyashita, K. Uchiyama, K. Satoh, N. Moeri, H. Okamoto, K. Miyagawa, K. Kanoda, M. Hedo, and Y. Uwatoko, J. Phys. IV \textbf{114}, 273 (2004).
\bibitem{005}D. J\'erome, Science \textbf{252}, 1509 (1991).
\bibitem{Mott1} K. Kanoda, Hyperfine Interactions \textbf{104}, 235 (1997).
\bibitem{Mott2} R. McKenzie, Science \textbf{278}, 820 (1997).

\bibitem{026}P. Garoche, R. Brusetti, D. J\'erome, K. Bechgaard, J. Phys. Lett. \textbf{43} 147 (1982).
\bibitem{027}N. Joo, P. Auban-Senzier, C.R. Pasquier, P. Monod, D. J\'erome, K. Bechgaard, EPJ B \textbf{40} 43 (2004).
\bibitem{028}M. Takigawa, H. Yasuoka, G. Saito, J. Phys. Soc. Jpn. \textbf{56} 873 (1987).
\bibitem{029}S. Belin, K. Behnia, Phys. Rev. Lett. \textbf{79} 2125 (1997).
\bibitem{006}I. J. Lee, S. E. Brown, W. G. Clark, M. J. Strouse, M. J. Naughton, W. Kang, and P. M. Chaikin, Phys. Rev. Lett. \textbf{88},  017004 (2002).
\bibitem{007}P. M. Chaikin, M.-Y. Choi, and R. L. Greene, J.Magn. Magn. Mater. \textbf{31}, 1268 (1983).
\bibitem{030}S. Yonezawa, Y. Maeno, K. Bechgaard, D. J\'erome, Phys. Rev. B \textbf{85}, 140502 (2012).
\bibitem{031}F. L. Pratt, T. Lancaster, S. J. Blundell, C. Baines, Phys. Rev. Lett. \textbf{110}, 107005 (2013).
\bibitem{032}H. Mayaffre, P. Wzietek, D. Jérome, C. Lenoir, P. Batail, Phys. Rev. Lett. \textbf{75}, 4122 (1995).
\bibitem{033}K. Kanoda, K. Miyagawa, A. Kawamoto, Y. Nakazawa, Phys. Rev. B \textbf{54}, 76 (1996).
\bibitem{034}S. M. De Soto, C. P. Slichter, A. M. Kini, H. H. Wang, U. Geiser, J. M. Williams, Phys. Rev. B \textbf{52}, 10364 (1995).
\bibitem{035}G. Koutroulakis, H. Kuhne, H.H. Wang, J.A. Schlueter, J. Wosnitza, S.E. Brown arXiv: \textbf{1601.06107}.
\bibitem{036}K. Izawa, H. Yamaguchi, T. Sasaki,  Y. Matsuda, Phys. Rev. Lett. \textbf{88}, 027002 (2001).
\bibitem{037}K. Ichimura,  K.  Nomura, Phys. Soc. Jpn. \textbf{75}, 051012 (2006).
\bibitem{038}T. Arai, K. Ichimura, K. Nomura, S. Takasaki, J. Yamada, S. Nakatsuji,  H. Anzai, Phys. Rev. B \textbf{63}, 104518 (2001).
\bibitem{039}K. Ichimura, M. Takami, K. Nomura, J. Phys. Soc. Jpn. \textbf{77}, 114707 (2008).
\bibitem{040}Y. Oka, H. Nobukane, N. Matsunaga, K. Nomura, J. Phys. Soc. Jpn. \textbf{84} 064713 (2015).
\bibitem{041}D. Guterding, S. Diehl, M. Altmeyer, T. Methfessel, U. Tutsch, H. Schubert, M. Lang, J. Muller, M. Huth, H. O. Jeschke, R. Valent\'i, M. Jourdan, H. Elmers, Phys. Rev. Lett. \textbf{116}, 237001 (2016).
\bibitem{042}D. Guterding, M. Altmeyer, H. O. Jeschke, and R. Valenti, Phys. Rev. B \textbf{94}, 024515 (2016).
\bibitem{043}H. Elsinger, J. Wosnitza, S. Wanka, J. Hagel, D. Schweitzer, W. Strunz, Phys. Rev. Lett. \textbf{84}, 6098 (2000).
\bibitem{044}J. Wosnitza, X. Liu, D. Schweitzer, H. J. Keller, Phys. Rev. B \textbf{50}, 12747 (1994).
\bibitem{045}S. Wanka, J. Hagel, D. Beckmann, J. Wosnitza, J. A. Schlueter, J. M. Williams, P. G. Nixon, R. W. Winter, G. L. Gard, Phys. Rev. B \textbf{57}, 3084 (1998).
\bibitem{046}J. Muller, M. Lang, R. Helfrich, F. Steglich, T. Sasaki, Phys. Rev. B \textbf{65}, 140509 (2002).
\bibitem{047}R. Beyer, J. Wosnitza, Low Temp. Phys. \textbf{39}, 225 (2013).

\bibitem{008}N. B. Hannay, T. H. Geballe, B. T. Matthias, K. Andres, P. Schmidt, and D. Macnair, Phys. Rev. Lett. \textbf{14}, 225 (1965).
\bibitem{009}A. F. Hebard, M. J. Rosseinsky, R. C. Haddon, D. W. Murphy, S. H. Glarum, T. T. M. Palstra, A. P. Ramirez, and A. R. Kortan, Nature \textbf{350}, 600 (1991).
\bibitem{010}K. Tanigaki, T. W. Ebbesen, S. Saito, J. Mizuki, J. S. Tsai, Y. Kubo, and S. Kuroshima, Nature \textbf{352}, 252 (1991).
\bibitem{011}T. T. M. Palstra, O. Zhou, Y. Iwasa, P. E. Sulewski, R. M. Fleming, and B. R. Zegarski, Solid State Commun. \textbf{93}, 327 (1995).
\bibitem{012}R. F. Kiefl, W. A. Macfarlane , K. H. Chow, S. Dunsiger, T. L. Duty, T. M. F. Johnston, J. W. Schneider, J. Sonier, L. Brard, R. M. Strongin, J. E. Fischer, and A. B. Smith, Phys. Rev. Lett. \textbf{70}, 3987 (1993).
\bibitem{013}O. Gunnarsson, Rev. Mod. Phys. \textbf{69}, 575 (1996).
\bibitem{014}R. Mitsuhashi, Y. Suzuki, Y. Yamanari, H. Mitamura, T. Kambe, N. Ikeda, H. Okamoto, A. Fujiwara, M. Yamaji, N. Kawasaki,Y. Maniwa, and Y. Kubozono, Nature \textbf{464}, 76 (2010).
\bibitem{015}X. F. Wang, R. H. Liu, Z. Gui, Y. L. Xie, Y. J. Yan, J. J. Ying, X. G. Luo, and X. H. Chen, Nat. Commun. \textbf{2}, 507 (2011).
\bibitem{016}M. Xue, T. Cao, D. Wang, Y. Wu, H. Yang, X. Dong, J. He, F. Li, and G. F. Chen, Sci. Rep. \textbf{2}, 389 (2012).
\bibitem{048}S. Heguri, M. Kobayashi, K. Tanigak\'i, Phys. Rev. B \textbf{92}, 014502 (2015).
\bibitem{049}Romero FD, Pitcher MJ, Hiley CI, Whitehead GFS, Kar S, Ganin AY, Antypov D, Collins C, Dyer MS, Klupp G, Colman RH, Prassides K, Rosseinsky MJ, Nat Chem. \textbf{7} 644 (2017).
\bibitem{017}R. S. Wang, Y. Gao, Z. B. Huang, and X. J. Chen, arXiv: \textbf{1703.06641}.
\bibitem{018}W. H. Liu, H. Lin, R. Z. Kang, X. Y. Zhu, Y. Zhang, S. Zheng, and H.-H. Wen, Phys. Rev. B \textbf{96}, 224501 (2017).
\bibitem{019}X. F. Wang, Y. J. Yan, Z. Gui, R. H. Liu, J. J. Ying, X. G. Luo, and X. H. Chen, Phys. Rev. B \textbf{84}, 214523 (2011).
\bibitem{020}Y. Kubozono, R. Eguchi, H. Goto, S. Hamao, T. Kambe, T. Terao, S. Nishiyama, L. Zheng, X. Miao, and H. Okamoto, J. Phys. Con. Mat. \textbf{28}, 334001 (2016).
\bibitem{VortexBaK122}H. Yang, B. Shen, Z. Wang, L. Shan, C. Ren, and H.-H. Wen, Phys. Rev. B \textbf{85}, 014524 (2012).

\bibitem{024}X. F. Wang, X. G. Luo, J. J. Ying, Z. J. Xiang, S. L. Zhang, R. R. Zhang, Y. H. Zhang, Y. G. Yan, A. F. Wang, P. Chen, G. J. Ye, X. H. Chen, J. Phys. Con. Mat. \textbf{24}, 345701 (2012).
\bibitem{025}G. A. Artioli, F. Hammerath, M. C. Mozzati, P. Carretta, F. Corana, B. Mannucci, S. Margadonna, L. Malavasi, Chem. Commun. \textbf{51} 1092 (2015).

\bibitem{021}H. Lei, X. Zhu, and C. Petrovic, EPL \textbf{95}, 17011 (2011).
\bibitem{022}Y. C. Lin and D. E. Williams, Acta Cryst. B \textbf{31}, 318 (1975).
\bibitem{023}J. Rodr\'iguez-Carvajal, Physica B \textbf{192}, 55 (1993).




\end{thebibliography}
\end{document}